\documentclass[journal]{IEEEtran}
\usepackage{amsmath, amssymb, bm, cite, epsfig, psfrag, mathtools}
\usepackage{epstopdf}
\usepackage{rotating}
\usepackage{dblfloatfix}
\usepackage{algorithm,algpseudocode}
\usepackage{array}
\usepackage{supertabular,booktabs}
\usepackage{ragged2e}
\newcolumntype{P}[1]{>{\centering\hspace{0pt}}p{#1}}
\newcolumntype{M}[1]{>{\centering\hspace{0pt}}m{#1}}
\newcolumntype{L}{>{\centering\arraybackslash}m{3cm}}
\usepackage{changepage}
\usepackage[font=small]{caption}
\usepackage{color,soul}
\usepackage{booktabs}

\usepackage{capt-of}
\usepackage{bbm}
\usepackage{multirow}
\usepackage[usenames,dvipsnames]{xcolor}
\renewcommand{\arraystretch}{1.5}

\usepackage{footnote}
\makesavenoteenv{tabular}
\makesavenoteenv{table}
\usepackage{etoolbox}
\usepackage{pbox}
\usepackage[hyphens]{url}
\usepackage[hidelinks]{hyperref}
\hypersetup{breaklinks=true}
\usepackage{cite}
\usepackage{longtable}
\usepackage{tabu}
\usepackage[top=0.45in,bottom=0.6in,left = 0.52in,right=0.52in]{geometry}

\usepackage{pdflscape}
\usepackage{fixltx2e}
\usepackage{graphicx}
\usepackage{tabu}
\usepackage{booktabs}
\def\PL{\textrm{PL}}
\def\dB{\textrm{dB}}
\def\FSPL{\textrm{FSPL}}
\def\CI{\textrm{CI}}
\def\CIF{\textrm{CIF}}
\def\CIX{\textrm{CIX}}
\def\CIFX{\textrm{CIFX}}
\def\FI{\textrm{FI}}
\def\ABG{\textrm{ABG}}
\def\ABGX{\textrm{ABGX}}

\def\XPD{\textrm{XPD}}
\def\1m{\textrm{1 m}}

\newtoggle{conference}
\togglefalse{conference} 
\usepackage{tikz}
\usetikzlibrary{calc}
\begin{document}
\bibliographystyle{IEEEtran}

\title{Indoor Office Plan Environment and Layout-Based MmWave Path Loss Models for 28 GHz and 73 GHz}

\author{
George R. MacCartney, Jr.,~\IEEEmembership{Student Member,~IEEE,}
Sijia Deng,~\IEEEmembership{Student Member,~IEEE,}\\
Theodore S. Rappaport,~\IEEEmembership{Fellow,~IEEE}

\thanks{This material is based upon work supported by the NYU WIRELESS Industrial Affiliates Program, three National Science Foundation (NSF) Research Grants: 1320472, 1302336, and 1555332, and the GAANN Fellowship Program. The authors wish to thank M. K. Samimi, K. Patade, and A. Hamza for their contribution to this project. G. R. MacCartney, Jr. (email: gmac@nyu.edu), S. Deng (email: sijia@nyu.edu), and T. S. Rappaport (email: tsr@nyu.edu), are with the NYU WIRELESS Research Center, NYU Tandon School of Engineering, Brooklyn, NY 11201.}
}
\maketitle
\begin{tikzpicture}[remember picture, overlay]
\node at ($(current page.north) + (0,-0.25in)$) {G. R. MacCartney, Jr., S. Deng, and T. S. Rappaport, ``Indoor Office Plan Environment and Layout-Based MmWave Path Loss Models for};
\node at ($(current page.north) + (0,-0.4in)$) {28 GHz and 73 GHz," to be published in \textit{2016 IEEE 83rd Vehicular Technology Conference Spring (VTC 2016-Spring)}, May 2016.};
\end{tikzpicture}
\begin{abstract}
This paper presents large-scale path loss models based on extensive ultra-wideband millimeter-wave propagation measurements performed at 28 GHz and 73 GHz in three typical indoor office layouts -- namely: corridor, open-plan, and closed-plan. A previous study combined all indoor layouts together, while this study separates them for site-specific indoor large-scale path loss model analysis. Measurements were conducted using a 400 megachips-per-second broadband sliding correlator channel sounder with 800 MHz first null-to-null RF bandwidth for 48 transmitter-receiver location combinations with distances ranging 3.9 m to 45.9 m for both co- and cross-polarized antenna configurations in line-of-sight and non-line-of-sight environments. Omnidirectional path loss values were synthesized from over 14,000 directional power delay profiles and were used to generate single-frequency and multi-frequency path loss models for combined, co-, and cross-polarized antennas. Large-scale path loss models that include a cross-polarization discrimination factor are provided for cross-polarized antenna measurements. The results show the value of using the close-in free space reference distance single and multi-frequency path loss models, as they offer simplicity (less parameters) in path loss calculation and prediction, without sacrificing accuracy. Moreover, the current 3GPP floating-intercept path loss model only requires a simple and subtle modification to convert to the close-in free space reference distance models. 
\end{abstract}

\iftoggle{conference}{}{
}

\section{Introduction}\label{sec:intro}
The overwhelming demand for broadband wireless communications is continuously growing and is expected to have an explosive increase in the next decade as the Internet-of-Things (IoT) expands and the use of multiple smart devices becomes commonplace~\cite{IoT:2014}. The limited availability of sub-6 GHz spectrum provides motivation for the use of millimeter-waves (mmWave) that contain a considerable amount of available raw bandwidth that could enable multi-gigabit-per-second communications for cellular, backhaul, office, and in-home applications as a part of fifth-generation (5G) wireless systems~\cite{PI:mmWave,WillWork:TSR13,TCOM15:RMSS,Niu2015Survey,SIGVTC2016}. MmWave bands (ranging from 30 GHz to 300 GHz) have been sparsely used for cellular or mobile applications, but recent research activities and government interest has developed regarding propagation characteristics and the feasibility of mmWave bands for 5G wireless broadband communications, such as 28 GHz (previously used for Local Multipoint Distribution Services (LMDS)), 60 GHz (expanding use for WiGig and Wireless-HD~\cite{Hansen:WiGig,WIGIG:WP}), and E-band (71-76 GHz and 81-86 GHz are lightly-licensed for point-to-point backhaul and vehicular applications), to name a few~\cite{PI:mmWave,TCOM15:RMSS}.

Indoor wireless services are currently administered over the 2.4 GHz and 5 GHz bands for WiFi, and at 60 GHz for WiGig. The vast available bandwidth (57-64 GHz) at 60 GHz (and unlicensed availability in the U.S. and other countries) motivated extensive 60 GHz indoor propagation measurements to understand and model channel characteristics necessary for designing indoor wireless local area networks (WLAN) capable of achieving multi-gigabits-per-second throughputs~\cite{Xu:60GHzJSAC,Malt:3}. The propagation studies at 60 GHz (fewer studies at other mmWave bands) focused on modeling path loss and multipath time dispersion in common office environments. 

Alvarez~\textit{et al}. studied the indoor radio channel between 1 and 9 GHz using omnidirectional antennas, and defined four environments and layouts which were line-of-sight (LOS) (direct path between the transmitter (TX) and receiver (RX)), Soft-NLOS (non-LOS) (no direct path, rather reflected paths between the TX and RX), Hard-NLOS (no direct or reflected paths between the TX and RX), and corridor (LOS case: direct path and many strong reflected paths)~\cite{Alvaro03}. The estimated path loss exponents (PLEs) relative to various reference distances were 1.4 ($d_0$ = 15.1 cm) for the LOS scenario (including corridor), 3.2 ($d_0$ = 8.2 cm) for Soft-NLOS, and 4.1 ($d_0$ = 6.7 cm) for Hard-NLOS. Geng~\textit{et al}. conducted indoor propagation measurement at 60 GHz in corridor, LOS hallway, and NLOS hallway environments, and the measured path loss attenuation slopes as a function of log-distance were 1.6 in LOS corridor, 2.2 in LOS hallway, and 3.0 in NLOS hallway environments~\cite{Geng60G}. Lei~\textit{et al}. investigated indoor LOS channel characteristics at 28 GHz, yielding a PLE of 2.2 for an open area (in-hall scenario), 1.8 for an office (smaller than the hall), and 1.2 for a corridor~\cite{Lei28G}. Zwick~\textit{et al.} also performed wideband channel measurements with omnidirectional TX and RX antennas in an indoor environment over 5 GHz of bandwidth, resulting in a PLE of 1.3 relative to 1 m free space path loss (FSPL) reference distance with a shadow factor of 5.1 dB~\cite{Zwick:1}. 

Path loss models are vital for understanding the attenuation of propagating signals and allow researchers and standards bodies to create accurate channel models for system-level network simulations that assist in the design of communications systems. Single frequency path loss models in most standards bodies are commonly presented in either the \emph{close-in free space reference distance (CI)} or \emph{floating intercept (FI)} form~\cite{PL5G:GMAC13,TCOM15:RMSS}. After investigating both models in~\cite{SD:WK15ICCIn}, single frequency and multi-frequency path loss models for separate and combined antenna polarizations were studied in~\cite{AccessIndoor:15} for a general indoor office environment at 28 GHz and 73 GHz. In this paper, a comprehensive study on layout-based single and multi-frequency mmWave path loss models are provided for separate and combined antenna polarizations in LOS and NLOS environments for three common indoor office layouts (corridor, open-plan, and closed-plan) at 28 GHz and 73 GHz. 

\section{Measurement Descriptions, Environments, and Layouts}\label{sec:MeasDesc}
\begin{figure*}[b!]
\centering
	\includegraphics[width=5.5in]{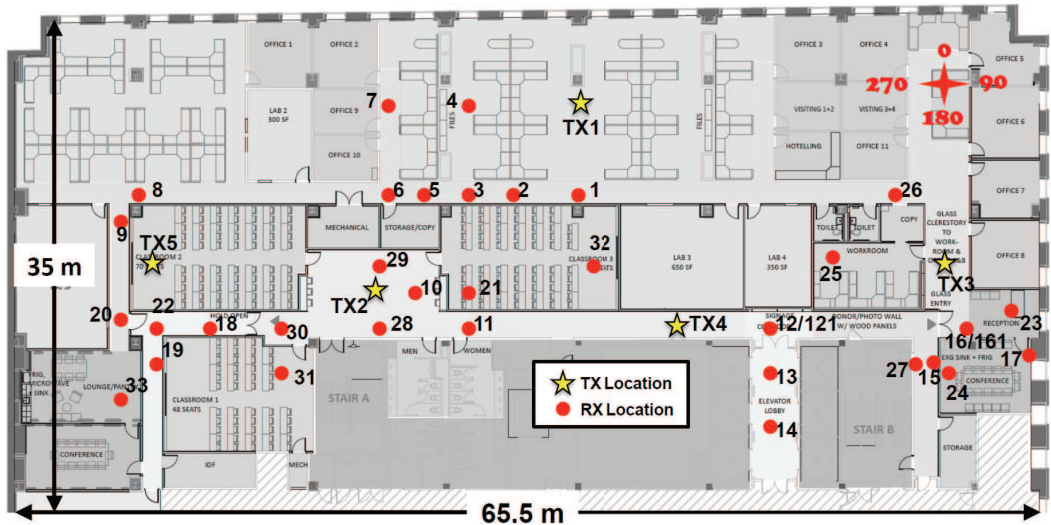}
	\caption{Map of the 2 MetroTech Center 9\textsuperscript{th} floor with five TX locations and 33 RX locations (some were used for multiple TX locations). Yellow stars represent the TX locations and red dots represent RX locations. The RX121 and RX161 locations were identical to the RX12 and RX16 locations, however the glass door near RX16 was propped open for RX121 and RX161 measurements, and was closed for RX12 and RX16 measurements. There were a total of 48 measured TX-RX location combinations.}
	\label{fig:MapAll}
\end{figure*}

The 28 GHz and 73 GHz propagation measurements were conducted in a typical office environment within a modern office building (35 m $\times$ 65.5 m) as shown in Fig.~\ref{fig:MapAll}, using a 400 Megachips-per-second (Mcps) spread spectrum broadband sliding correlator channel sounder and a pair of mechanically-steerable highly-directional horn antennas at the TX and RX (15 dBi gain, 30$^{\circ}$ azimuth half-power beamwidth (HPBW) at 28 GHz and 20 dBi, 15$^{\circ}$ azimuth HPBW at 73 GHz)~\cite{SD:WK15ICCIn,AccessIndoor:15}. Measurements at both frequencies were conducted at the same five TX and 33 RX locations with transmitter-receiver (T-R) separation distances ranging from 3.9 m to 45.9 m, for a total of 48 TX-RX location combinations (10 in LOS and 38 in NLOS) with the TX and RX antennas at 2.5 m and 1.5 m heights, respectively, to emulate an indoor hotspot scenario.

Three common indoor office layouts were measured: corridor, open-plan, and closed-plan, with layout descriptions provided in Table~\ref{tbl:EnvDesc} and the corresponding TX-RX combination environments and layouts given in Table~\ref{tbl:TXRXCom}. For each measured TX-RX location combination and polarization, eight unique antenna pointing angle measurement sweeps were performed at both the TX and RX to investigate angle of departure (AOD) and angle of arrival (AOA) statistics for different elevation planes, in order to synthesize omnidirectional path loss from directional measurements as described in~\cite{AccessIndoor:15,TCOM15:RMSS}. The measurements included vertical-to-vertical (V-V) and vertical-to-horizontal (V-H) antenna polarization configurations between the TX and RX to study the impact of polarization on indoor mmWave propagation. Detailed hardware descriptions and measurement procedures can be found in \cite{AccessIndoor:15}. The measurements were taken in a semi-static environment with little if any moving objects, thus the models to follow are useful for large-scale path loss, while additional work with channel dynamics is on-going.

\begin{table}
	\centering
	\renewcommand{\arraystretch}{2}
	\caption{Indoor office layout descriptions.} \label{tbl:EnvDesc}
	\fontsize{8}{8}\selectfont
	\begin{tabular}{ | c ||p{6cm} |}
	\hline
	\textbf{Layout} & \textbf{Description}\\ \hline \hline \cline{1-2}
	\textbf{Corridor} & A narrow corridor hallway in which the propagating signal travels down a corridor to reach the RX by a LOS path, reflections, and/or diffraction, but \textbf{not} penetration.\\ \hline
	\textbf{Open-Plan} & A typical office-space with a large layout and central TX location, where the propagating signal reaches the RX by a LOS path, reflections, and/or diffraction, but \textbf{not} penetration.\\ \hline
	\textbf{Closed-Plan} & A typical office-space where the propagating signal must penetrate an obstruction to reach the RX in addition to potential reflections, and/or diffraction.\\ \hline
	\end{tabular}
\end{table}

\begin{table}
	\caption{Corresponding TX and RX locations measured for each layout and environment (LOS or NLOS). ``CO" stands for corridor, ``OP" stands for open-plan, ``CP" stands for closed-plan, and ``Scn." stand for scenario. A ``-" indicates that no TX-RX location combination was measured for the specified layout and environment. Refer to the map in Fig.~\ref{fig:MapAll}.} \label{tbl:TXRXCom}
	\fontsize{6}{6}\selectfont
	\begin{center}
		\begin{tabular}{|M{0.6cm}|M{0.6cm}|M{0.9cm}|M{0.9cm}|M{0.9cm}|M{1.1cm}|M{0.9cm}|}
		\hline
		\multirow{2}{*}{\bf{Scn.}} & \multirow{2}{*}{\bf{Env.}} & \multicolumn{5}{c|}{\textbf{TX IDs}} \\ \cline{3-7}
		& & \bf{1} & \bf{2} & \bf{3} & \bf{4} & \bf{5} \tabularnewline \hline
		\multirow{4}{*}{\bf{CO}} & \bf{LOS} & - & - & - & RX11, 12, 28, 121, 161 & - \tabularnewline \cline{2-7}
							& \bf{NLOS} & - & RX11, 12, 15, 161 & - & RX15 & - \tabularnewline \hline
		\multirow{4}{*}{\bf{OP}} & \bf{LOS}& RX1, 4, 7 & RX10 & RX16 & - & - \tabularnewline \cline{2-7}
							& \bf{NLOS} & RX2, 3, 5, 6, 8, 9 & - & RX23, 26 & - & - \tabularnewline \hline
		\bf{CP} 			& \bf{NLOS} & - & RX13, 14, 16--22 & RX17, 24, 25, 27 & RX13, 14, 16, 18 & RX8, 19, 28-33 \tabularnewline \hline
		\end{tabular}
	\end{center}
\end{table}

\section{Path Loss Model Definitions}
The CI path loss model has been used for decades to describe path loss relative to path loss at a close-in free space reference distance (in view of the transmitter; 1 km for early urban-macro (UMa) models such as the Hata-model, and more recently a 1 m distance for mmWave frequencies and urban-micro (UMi) scenarios~\cite{Zwick:1,TCOM15:RMSS}). The CI model is defined by the PLE $n$:
\begin{equation}\label{eq:CI}
\begin{split}
\PL^{\CI}(f,d)[\dB]=\FSPL(f,d_0)+10n\log_{10}\left(\frac{d}{d_0}\right)+X_{\sigma}^{\CI}\\ \mathrm{for}\;\; d\geq d_0\text{,   with } d_0=1\textrm{ m}
\end{split}
\end{equation}
where $X_{\sigma}^{\CI}$ is a zero mean Gaussian random variable with standard deviation $\sigma$ in dB, which models shadow fading (SF)~\cite{Rappaport:Wireless2nd}. The CI model uses a physically-based reference distance $d_0$, where $\FSPL(f,d_0) = 10\log_{10}\left(\frac{4\pi d_0f}{c}\right)^2$, $f$ is the carrier frequency and $c$ is the speed of light. The PLE $n$ is found via minimum mean square error (MMSE) and the closed-form expression for optimizing $n$ is given in~\cite{AccessIndoor:15}, along with closed-form parameter optimizations for all path loss models presented in this paper. 

The \emph{close-in free space reference distance with cross-polarization discrimination (XPD) factor (CIX)} model is an extension of the co-polarized CI model that has an added XPD term in dB given by:
\begin{equation}\label{eq:CIX}
\begin{split}
\PL^{\CIX}(f,d)[\dB]=\FSPL(f,d_0)+10n_{\text{(V-V)}}\log_{10}\left(\frac{d}{d_0}\right)\\+\textrm{XPD}[\dB]+X_{\sigma}^{\CIX}
\end{split}
\end{equation}
where $n_{\text{(V-V)}}$ is the co-polarized PLE from the CI model. The XPD factor corresponds to a constant optimized attenuation term that minimizes the error between the estimated model and cross-polarized measurements while using the CI co-polarized PLE. 

The FI path loss model is used in the 3GPP and WINNER II standards~\cite{TCOM15:RMSS}. This two parameter model does not include a physically-based leverage point, although it is a least-squares best-fit estimator to the measured data, and has a similar form to~\eqref{eq:CI}: 
\begin{equation}\label{eq:FI}
\PL^{\FI}(d)[\dB] = \alpha+10\cdot\beta\log_{10}(d)+X^{\FI}_{\sigma}
\end{equation}
where $\alpha$ is the floating-intercept in dB (different than a FSPL reference), and $\beta$ is the slope of the line (different than a PLE) with a large-scale shadowing random variable $X^{\FI}_{\sigma}$. Previous studies showed that the CI and FI path loss models perform similarly over identical datasets, with shadow fading standard deviations that are within a fraction of a dB of each other~\cite{TCOM15:RMSS,Sulyman:ComMag14,PL5G:GMAC13}.

Standards bodies and modeling groups are also interested in multi-frequency path loss models in order to have a general form/model to cover a broad range of frequencies and measurements. The CI model~\eqref{eq:CI} can be used for both single and multi-frequency datasets. The \emph{alpha-beta-gamma (ABG)} model is a common multi-frequency model that uses a frequency-dependent and distance-dependent term to estimate path loss~\cite{PL5G:GMAC13,ABG:2012}:
\begin{equation}\label{eq:ABG}
\begin{split}
\PL^{\ABG}(f,d)[\dB]=10\alpha\log_{10}\left(\frac{d}{d_0}\right)+\beta\\+10\gamma\log_{10}\left(\frac{f}{\text{1 GHz}}\right)+X^{\ABG}_{\sigma}\text{,   with } d_0=1\textrm{ m}
\end{split}
\end{equation}
where $\alpha$ and $\gamma$ are coefficients that describe the distance and frequency dependency on path loss, while $\beta$ is the optimized offset in path loss, $f$ is the carrier frequency in GHz, and $X^{\ABG}_{\sigma}$ is a zero mean Gaussian SF random variable with standard deviation $\sigma$. The ABG model is an extension of the FI model for multiple frequencies and is solved via MMSE to minimize $\sigma$ by simultaneously solving for $\alpha$, $\beta$, and $\gamma$. The closed-form expressions that optimize the ABG model parameters are provided in~\cite{AccessIndoor:15}.

The ABG model extends to an \emph{ABGX} model (similar to CIX) for cross-polarized antenna measurements:
\begin{equation}\label{eq:ABGX}
\begin{split}
\PL^{\ABGX}(f,d)[\dB]=10\alpha\log_{10}\left(\frac{d}{d_0}\right)+\beta\\+10\gamma\log_{10}\left(\frac{f}{\text{1 GHz}}\right)+\XPD[\dB]+X^{\ABGX}_{\sigma}\text{,   with } d_0=1\textrm{ m}
\end{split}
\end{equation}
where the $\alpha$, $\beta$, and $\gamma$ values determined for the ABG model are used to solve for the optimal XPD factor that minimizes $\sigma$ via MMSE. 

A relatively new multi-frequency model first presented in~\cite{AccessIndoor:15,Shu:VTC2016} is an extension of the CI model with a frequency dependent term, the \emph{close-in free space reference distance with frequency weighting (CIF)} path loss model. The CIF model also includes a 1 m FSPL anchoring point like the CI model:
\begin{equation}\label{eq:CIF}
\begin{split}
\PL^{\CIF}(f,d)[\dB]=\FSPL(f,d_0)+\\10n\Bigg(1+b\left(\frac{f-f_0}{f_0}\right)\Bigg)\log_{10}\left(\frac{d}{d_0}\right)+X^{\CIF}_{\sigma}\text{,   with } d_0=1\textrm{ m}
\end{split}
\end{equation}
where $n$ denotes the distance dependency of path loss, $b$ is a model-fitting parameter that captures the linear frequency dependency of path loss, $f_0$ is a fixed reference frequency that balances the linear frequency dependence, and is based on the average of all frequencies represented from the measured datasets~\cite{AccessIndoor:15}, and $X^{\CIF}_{\sigma}$ is a zero mean Gaussian random variable (in dB) that describes shadow fading. The $f_0$ values for the five different layouts and environments are provided in Table~\ref{tbl:OmniMFPL}.

The CIF model also extends to a CIFX model and uses the CIF co-polarized model parameters to determine the optimal XPD factor that minimizes $\sigma$ for cross-polarized measurements:
\begin{equation}\label{eq:CIFX}
\begin{split}
\PL^{\CIFX}(f,d)[\dB]=\FSPL(f,d_0)+\\10n\Bigg(1+b\left(\frac{f-f_0}{f_0}\right)\Bigg)\log_{10}\left(\frac{d}{d_0}\right)\\+\textrm{XPD}[\dB]+X^{\CIFX}_{\sigma}\text{,   with } d_0=1\textrm{ m}
\end{split}
\end{equation}
where the $n$ and $b$ values found from the CIF model and the same $f_0$ parameter are used to solve for the XPD factor via MMSE. The closed-form expression that optimizes the XPD factor for the CIFX model is given in~\cite{AccessIndoor:15,Shu:VTC2016}. 

The optimized minimum error CI PLE parameter (see Appendix in~\cite{AccessIndoor:15}) is found by first subtracting the 1 m FSPL value from each path loss data point to solve for $n$~\cite{Rappaport:mmWave,TCOM15:RMSS,Thomas16a}. The CI model can then be applied across a broad range of frequencies with~\eqref{eq:CI} by using the single value of $n$. The ABG model also is applied across a broad range of frequencies, although with three model parameters. However, as shown in~\cite{TCOM15:RMSS,Thomas16a} the floating parameters (for FI or ABG) vary sporadically across different frequencies, meaning the ABG model is more prone to error when extrapolating the model outside of the frequencies or distances from which data was used to optimize the parameters. The CI model PLE, however, remains stable across a broad range of frequencies, distances, and environments~\cite{Thomas16a}.

\section{Omnidirectional Path Loss Model Parameters}
Omnidirectional path loss data was synthesized from measurements using directional antennas where PDPs were recorded at unique antenna pointing angles between the TX and RX over numerous azimuth and elevation angle positions as described in~\cite{TCOM15:RMSS,AccessIndoor:15}. 

\subsection{Single Frequency Models}
\subsubsection{Single Frequency Models for Co- and Cross-Polarized Antennas}
\begin{figure}[b!]
	\centering
	\includegraphics[width=3.7in]{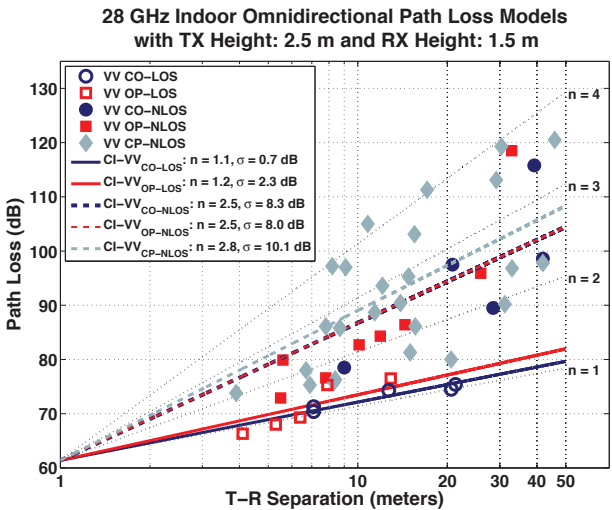}
	\caption{28 GHz omnidirectional scatter plot and CI path loss model parameters for LOS and NLOS environments for corridor (CO), open-plan (OP), and closed-plan (CP) indoor layouts.}
	\label{fig:28CI}
\end{figure}
Fig.~\ref{fig:28CI} displays the 28 GHz CI ($d_0$ =  1 m) omnidirectional path data and models for five LOS and NLOS indoor office environments and layouts for V-V polarized antennas, and Table~\ref{tbl:OmniSFPL} summarizes the single frequency CI and FI model parameters for separate and combined antenna polarizations. Fig.~\ref{fig:28CI} shows similarities between 28 GHz path losses in the LOS corridor and LOS open-plan layouts with PLEs of 1.1 and 1.2, respectively, which are significantly smaller than theoretical free space path loss (n = 2), due to constructive interference and waveguiding effects. The NLOS corridor and open-plan layouts have identical PLEs of 2.5, whereas the closed-plan layout results in a higher PLE of 2.8 at 28 GHz, due to high penetration loss caused by obstructions. The NLOS corridor and open-plan layouts at 73 GHz also have identical PLEs of 3.1, while the PLE is 3.3 for closed-plan. The higher PLEs at 73 GHz compared to 28 GHz for all environments and layouts are the result of increased scattering and attenuation due to the smaller wavelengths at higher frequencies. 

\begin{table}
	\centering
	\renewcommand{\arraystretch}{1.3}
	\caption{Single frequency omnidirectional CI and FI path loss model parameters for 28 GHz and 73 GHz for combined (Comb.), co (V-V), and cross (V-H) polarized antennas for LOS and NLOS environments. ``Freq." stands for frequency, ``Pol." stands for TX-RX antenna polarization, ``Comb." stands for combined TX-RX antenna polarization, ``Env." stands for environment. ``L/O." stands for layout, ``co" stands for corridor, ``op" stands for open-plan, and ``cp" stands for closed-plan. $\Delta\sigma = \sigma_{CI} - \sigma_{FI}$, is the difference in SF standard deviation between CI and FI models.} \label{tbl:OmniSFPL}
	\fontsize{8}{8}\selectfont
	\scalebox{0.78}{
	\begin{tabu}{|c|c|c|c|c|c|[1.5pt]c|c|c|[1.5pt]c|}  \hline
	\multicolumn{10}{|c|}{\textbf{Single Frequency Omnidirectional CI and FI Path Loss Models}} \\ 
	\multicolumn{10}{|c|}{\textbf{For Separate and Combined Antenna Polarizations}} \\ \specialrule{1.5pt}{0pt}{0pt}
	\multirow{2}{*}{\bf{Freq.}} & \multirow{2}{*}{\bf{Pol.}} &\multirow{2}{*}{\bf{Env.}} & \multirow{2}{*}{\bf{L/O}} & \multicolumn{2}{c| [1.5pt]}{\textbf{CI: {$\bm{d_0} = 1$ m}}} & \multicolumn{3}{c| [1.5pt]}{\textbf{FI}} & $\bm{\Delta\sigma}$  \\ \cline{5-9}
	& & & & \bf{PLE} & \bf{$\bm{\sigma}$ [dB]} & \bf{$\bm{\alpha}$ [dB]} & $\bm{\beta}$ & \bf{$\bm{\sigma}$ [dB]} & [dB] \\ \specialrule{1.5pt}{0pt}{0pt}
	\multirow{15}{*}{28 GHz}	& \multirow{5}{*}{V-V}	& \multirow{2}{*}{LOS}		& co	& 1.1	& 0.7	& 63.6	& 0.9	& 0.6 	& 0.1		\\ \cline{4-10}
	&							&													& op	& 1.2	& 2.3	& 52.3	& 2.3	& 1.4	& 0.9		\\ \cline{3-10}
	&													&  \multirow{3}{*}{NLOS}	& co	& 2.5	& 8.3	& 40.7	& 4.0	& 7.5	& 0.8		\\ \cline{4-10}
	&							&													& op	& 2.5	& 8.0	& 38.5	& 4.6	& 5.7	& 2.3		\\ \cline{4-10}
	&							&													& cp	& 2.8	& 10.1	& 55.0	& 3.3	& 10.0	& 0.1		\\ \cline{2-10}
								& \multirow{5}{*}{V-H}	& \multirow{2}{*}{LOS}		& co	& 2.4	& 2.8	& 76.1	& 1.1	& 0.2	& 2.6		\\ \cline{4-10}
	&							&													& op	& 2.8	& 1.6	& 66.7	& 2.2	& 1.1	& 0.5		\\ \cline{3-10}
	&													& \multirow{3}{*}{NLOS}		& co	& 3.2	& 3.3	& 58.4	& 3.4	& 3.3	& 0		    \\ \cline{4-10}
	&							&													& op	& 3.4	& 4.0	& 58.7	& 3.7	& 3.9	& 0.1		\\ \cline{4-10}
	&							&													& cp	& 3.7	& 10.7	& 61.3	& 3.8	& 10.7	& 0			\\ \cline{2-10}
								& \multirow{5}{*}{Comb.}& \multirow{2}{*}{LOS}		& co	& 1.7	& 7.4	& 69.8	& 1.0	& 7.3	& 0.1		\\ \cline{4-10}
	&							&													& op	& 2.0	& 6.9	& 59.5	& 2.2	& 6.9	& 0			\\ \cline{3-10}
	&													& \multirow{3}{*}{NLOS}		& co	& 2.8	& 8.0	& 51.5	& 3.5	& 7.8	& 0.2		\\ \cline{4-10}
	&							&													& op	& 2.9	& 7.9	& 50.2	& 3.9	& 7.5	& 0.4		\\ \cline{4-10}
	&							&													& cp	& 3.2	& 11.8	& 59.2	& 3.4	& 11.8	& 0			\\ \specialrule{1.5pt}{0pt}{0pt}
	\multirow{15}{*}{73 GHz}	& \multirow{5}{*}{V-V}	& \multirow{2}{*}{LOS}		& co	& 1.2	& 2.3	& 81.4	& 0.2	& 0.8	& 1.5		\\ \cline{4-10}
	&							&													& op	& 1.5	& 1.3	& 72.5	& 1.2	& 1.2	& 0.1		\\ \cline{3-10}
	&													& \multirow{3}{*}{NLOS}		& co	& 3.1	& 13.4	& 51.2	& 4.4	& 13.1	& 0.3		\\ \cline{4-10}
	&							&													& op	& 3.1	& 6.8	& 66.9	& 3.4	& 6.8	& 0			\\ \cline{4-10}
	&							&													& cp	& 3.3	& 11.7	& 82.6	& 2.2	& 11.4	& 0.3		\\ \cline{2-10}
								& \multirow{5}{*}{V-H}	& \multirow{2}{*}{LOS}		& co	& 3.3	& 5.9	& 100.5	& 0.6	& 1.2	& 4.7		\\ \cline{4-10}
	&							&													& op	& 4.0	& 4.5	& 88.5	& 1.8	& 2.5	& 2.0		\\ \cline{3-10}
	&													& \multirow{3}{*}{NLOS}		& co	& 4.0	& 7.5	& 92.7	& 2.3	& 6.3	& \textcolor{black}{1.2}		\\ \cline{4-10}
	&							&													& op	& 4.4	& 6.8	& 99.8	& 1.3	& 4.7	& 2.1		\\ \cline{4-10}
	&							&													& cp	& 4.7	& 10.0	& 99.4	& 2.1	& 7.5	& 2.5		\\ \cline{2-10} 
								& \multirow{5}{*}{Comb.}& \multirow{2}{*}{LOS}		& co	& 2.2	& 12.4	& 91.0	& 0.4	& 11.7	& 0.7		\\ \cline{4-10}
	&							&													& op	& 2.8	& 11.1	& 80.5	& 1.5	& 10.9	& 0.2		\\ \cline{3-10}
	&													& \multirow{3}{*}{NLOS}		& co	& 3.5	& 12.8	& 74.0	& 3.2	& 12.8	& 0			\\ \cline{4-10}
	&							&													& op	& 3.6	& 9.3	& 84.6	& 2.2	& 8.8	& 0.5		\\ \cline{4-10}
	&							&													& cp	& 4.0	& 13.5	& 92.9	& 2.0	& 12.4	& 1.1		\\ \specialrule{1.5pt}{0pt}{0pt}
	\end{tabu}}
\end{table}

\begin{figure}[b!]
	\centering
	\includegraphics[width=3.7in]{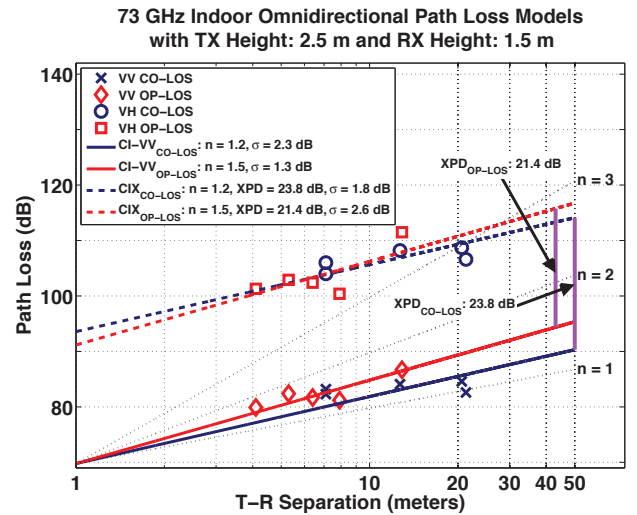}
	\caption{73 GHz omnidirectional CI and CIX path loss model scatter plots and parameters for LOS corridor (CO) and LOS open-plan (OP) indoor layouts, displaying the XPD factors.}
	\label{fig:73CIX}
\end{figure}

Table~\ref{tbl:OmniSFXPDPL} provides the CIX model parameters for 28 GHz and 73 GHz cross-polarized measurements. The CIX models resulted in larger XPD factors in LOS than in NLOS, especially the corridor and open-plan layouts which resulted in XPD factors of 14.6 dB and 13.3 dB in LOS, and 8.8 dB and 8.7 dB in NLOS environments at 28 GHz, respectively. Fig.~\ref{fig:73CIX} shows the 73 GHz CI and CIX path loss models in LOS corridor and open-plan layouts and the corresponding XPD factors. The 73 GHz XPD factors are 23.8 dB and 21.4 dB for the LOS corridor and open-plan layouts, respectively, indicating large cross-polarization isolation in LOS indoor office environments at 73 GHz and the potential for dual-polarized indoor communications systems at mmWave. 

The two-parameter FI model provides extremely low LOS $\beta$ slope values of 0.2 and 0.4 in LOS corridors for V-V and combined polarization antennas at 73 GHz, respectively, showing the sensitivity and lack of physical interpretation of the FI model. The CI model, however, uses only one parameter based on the 1 m free space reference distance, allowing for simple calculations and easy prediction of path loss values. From the comparison of the SF standard deviations for CI and FI models, it is apparent that for V-V antennas, the SF standard deviations for the CI and FI models differ by only a fraction of a dB in most cases at 28 GHz and 73 GHz (with the exception of a 2.3 dB difference in the NLOS open-plan layout at 28 GHz and a 1.5 dB difference in the LOS corridor layout at 73 GHz). The CIX models, by introducing an XPD factor, result in lower SF standard deviations compared to CI models for cross-polarized antenna scenarios (V-H)~\cite{AccessIndoor:15}, and are within 1 dB of FI model standard deviations, indicating the value of using CIX models for cross-polarized antennas measurements.

\begin{table}
	\centering
	\caption{Single frequency omnidirectional CIX path loss model parameters with $d_0 = 1$ m for 28 GHz and 73 GHz indoor channels for cross-polarized antennas (V-H) for LOS and NLOS environments and various indoor layouts. ``Freq." stands for carrier frequency, ``Pol." stands for TX-RX antenna polarization, ``Env." stands for environment. ``L/O." stands for layout. ``co" stands for corridor, ``op" stands for open-plan, and ``cp" stands for closed-plan.} \label{tbl:OmniSFXPDPL}
	\fontsize{8}{8}\selectfont
	\scalebox{0.9}{
	\begin{tabu}{|c|c|c|c|c|c|c|}  \hline
	\multicolumn{7}{|c|}{\textbf{Single Frequency Omnidirectional CIX Path Loss Models}} \\ \specialrule{1.5pt}{0pt}{0pt}
	\multirow{2}{*}{\bf{Freq.}} & \multirow{2}{*}{\bf{Pol.}} &\multirow{2}{*}{\bf{Env.}} & \multirow{2}{*}{\bf{L/O}} & \multicolumn{3}{c|}{\textbf{CIX: {$\bm{d_0} = 1$ m}}}  \\ \cline{5-7}
	& & & & \bf{$\bm{n}_{(\text{V-V})}$} & \bf{XPD [dB]} & \bf{$\bm{\sigma}$ [dB]}  				\\ \specialrule{1.5pt}{0pt}{0pt}
	\multirow{5}{*}{28 GHz}	& \multirow{5}{*}{V-H}	& \multirow{2}{*}{LOS}	& co	& 1.1	& 14.6	& 0.2	\\ \cline{4-7}
	&												&						& op	& 1.2	& 13.3	& 2.0	\\ \cline{3-7}
	&												& \multirow{3}{*}{NLOS}	& co	& 2.5	& 8.8	& 3.9	\\ \cline{4-7}
	&												&						& op	& 2.5	& 8.7	& 4.7	\\ \cline{4-7}
	&												&						& cp	& 2.8	& 11.0	& 11.0	\\ \specialrule{1.5pt}{0pt}{0pt}
	\multirow{5}{*}{73 GHz}	& \multirow{5}{*}{V-H}	& \multirow{2}{*}{LOS}	& co	& 1.2	& 23.8	& 1.8	\\ \cline{4-7}
	&												&						& op	& 1.5	& 21.4	& 2.6	\\ \cline{3-7}
	&												& \multirow{3}{*}{NLOS}	& co	& 3.1	& 12.9	& 6.5	\\ \cline{4-7}
	&												&						& op	& 3.1	& 12.9	& 5.5	\\ \cline{4-7}
	&												&						& cp	& 3.3	& 16.5	& 8.1	\\ \specialrule{1.5pt}{0pt}{0pt}
	\end{tabu}}
\end{table}

\subsubsection{Single Frequency Models for Combined Antenna Polarization}
In order to characterize path loss regardless of polarization, the co- and cross-polarization measurements were lumped into one dataset for 28 GHz and 73 GHz to generate path loss models that may be applied for arbitrary antenna polarization situations (commonly used to model the various orientations of a mobile hand-set). Table~\ref{tbl:OmniSFPL} provides single frequency CI and FI path loss model parameters for combined antenna polarizations at 28 GHz and 73 GHz. The 28 GHz CI models resulted in smaller PLEs and SF standard deviations in all measured environments and layouts compared to the corresponding PLEs and SF standard deviations for 73 GHz, indicating larger attenuation and more signal level variability at higher frequencies. It is worth noting that SF standard deviation differences for the CI and FI models are mostly within 1 dB for all measured environments and layouts, and the CI model requires only one parameter with a physically-based FSPL anchoring point. 
\begin{table}
	\centering
	\renewcommand{\arraystretch}{1.3}
	\caption{Multi-frequency omnidirectional path loss model parameters in LOS and NLOS environments and layouts. The CIX, CIFX, and ABGX cross-polarized models use the parameters found for their respective co-polarized models to determine the XPD factor that minimizes $\sigma$. ``Pol." stands for TX-RX antenna polarization.}\label{tbl:OmniMFPL}
	\scalebox{0.68}{
		\fontsize{8}{8}\selectfont
		\begin{tabular}{|c|c|c|c|c|c|c|} \specialrule{1.5pt}{0pt}{0pt}
			\multicolumn{7}{|c|}{\textbf{28 GHz and 73 GHz Multi-Frequency Omnidirectional}} 						\\
			\multicolumn{7}{|c|}{\textbf{LOS Corridor Path Loss Model Parameters}} 									\\ \specialrule{1.5pt}{0pt}{0pt}
			&  		Pol. 	& PLE 		& \multicolumn{2}{c|}{XPD [dB]} 		& \multicolumn{2}{c|}{$\sigma$ [dB]}\\ \hline
			CI 		& V-V 	& 1.1 		& \multicolumn{2}{c|}{-} 				& \multicolumn{2}{c|}{1.9 } 		\\ \hline
			CIX 	& V-H 	& 1.1 		& \multicolumn{2}{c|}{19.2} 			& \multicolumn{2}{c|}{5.5 } 		\\ \specialrule{1.5pt}{0pt}{0pt}
			& 		Pol. 	& $n$ 		& $b$ 		& $f_0$ [GHz] 	& XPD [dB]	& $\sigma$ [dB] 					\\ \hline
			CIF 	& V-V 	& 1.1 		& 0.13 		& 51  		& - 		& 1.7  									\\ \hline
			CIFX	& V-H 	& 1.1 		& 0.13 		& 51  		& 19.2 		& 4.8 									\\ \specialrule{1.5pt}{0pt}{0pt}
			& 		Pol. 	& $\alpha$ 	& $\beta$ 	& $\gamma$ 		& XPD [dB]  & $\sigma$ [dB] 					\\ \hline
			ABG 	& V-V  	& 0.5 		& 32.2 		& 2.4 			& - 		& 1.0  								\\ \hline
			ABGX	& V-H 	& 0.5 		& 32.2 		& 2.4  			& 18.9  	& 4.6  								\\ \specialrule{1.5pt}{0pt}{0pt}
			
			\multicolumn{7}{|c|}{\textbf{28 GHz and 73 GHz Multi-Frequency Omnidirectional}} 						\\
			\multicolumn{7}{|c|}{\textbf{LOS Open-Plan Path Loss Model Parameters}} 								\\ \specialrule{1.5pt}{0pt}{0pt}
			& 		Pol. 	& PLE 		& \multicolumn{2}{c|}{XPD [dB]} 		& \multicolumn{2}{c|}{$\sigma$ [dB]}\\ \hline
			CI 		& V-V 	& 1.4 		& \multicolumn{2}{c|}{-} 				& \multicolumn{2}{c|}{2.2 } 		\\ \hline
			CIX 	& V-H 	& 1.4 		& \multicolumn{2}{c|}{17.3} 			& \multicolumn{2}{c|}{5.8 } 		\\ \specialrule{1.5pt}{0pt}{0pt} 
			& 		Pol. 	& $n$ 		& $b$ 		& $f_0$ [GHz]	& XPD [dB]	& $\sigma$ [dB]						\\ \hline
			CIF 	& V-V 	& 1.4 		& 0.24 		& 51 			& - 		& 1.9  								\\ \hline
			CIFX 	& V-H 	& 1.4 		& 0.24 		& 51 		 	& 17.3  	& 4.7 								\\ \specialrule{1.5pt}{0pt}{0pt} 
			& 		Pol. 	& $\alpha$ 	& $\beta$ 	& $\gamma$ 		& XPD [dB]	& $\sigma$ [dB]						\\ \hline
			ABG 	& V-V 	& 1.7 		& 17.8 		& 2.7  			& - 		& 1.6  								\\ \hline
			ABGX 	& V-H 	& 1.7 		& 17.8 		& 2.7  			& 17.5  	& 4.4  								\\ \specialrule{1.5pt}{0pt}{0pt}
			
			\multicolumn{7}{|c|}{\textbf{28 GHz and 73 GHz Multi-Frequency Omnidirectional}} 						\\ 
			\multicolumn{7}{|c|}{\textbf{NLOS Corridor Path Loss Model Parameters}} 								\\ \specialrule{1.5pt}{0pt}{0pt}
			& 		Pol. 	& PLE 		& \multicolumn{2}{c|}{XPD [dB]} 		& \multicolumn{2}{c|}{$\sigma$ [dB]}\\ \hline
			CI 		& V-V 	& 2.8 		& \multicolumn{2}{c|}{-} 				& \multicolumn{2}{c|}{11.8 } 		\\ \hline
			CIX 	& V-H 	& 2.8 		& \multicolumn{2}{c|}{10.8 } 			& \multicolumn{2}{c|}{7.7 } 		\\ \specialrule{1.5pt}{0pt}{0pt}
			& 		Pol. 	& $n$ 		& $b$ 		& $f_0$ [GHz]	& XPD [dB]	& $\sigma$ [dB]						\\ \hline
			CIF 	& V-V 	& 2.8 		& 0.22 		& 51  			& - 		& 11.2  							\\ \hline
			CIFX 	& V-H 	& 2.8 		& 0.22 		& 51  			& 10.8  	& 5.8  								\\ \specialrule{1.5pt}{0pt}{0pt} 
			& 		Pol. 	& $\alpha$ 	& $\beta$ 	& $\gamma$ 		& XPD [dB]  & $\sigma$ [dB]						\\ \hline
			ABG 	& V-V 	& 4.2 		& -17.2 	& 3.8  			& - 		& 10.7  							\\ \hline
			ABGX 	& V-H 	& 4.2 		& -17.2 	& 3.8  			& 12.1  	& 6.4  								\\ \specialrule{1.5pt}{0pt}{0pt}
			
			\multicolumn{7}{|c|}{\textbf{28 GHz and 73 GHz Multi-Frequency Omnidirectional}} 						\\ 
			\multicolumn{7}{|c|}{\textbf{NLOS Open-Plan Path Loss Model Parameters}} 								\\ \specialrule{1.5pt}{0pt}{0pt}
			& 		Pol. 	& PLE 		& \multicolumn{2}{c|}{XPD [dB]} 		& \multicolumn{2}{c|}{$\sigma$ [dB]}\\ \hline
			CI 		& V-V 	& 2.8 		& \multicolumn{2}{c|}{-} 				& \multicolumn{2}{c|}{8.0 } 		\\ \hline
			CIX 	& V-H 	& 2.8 		& \multicolumn{2}{c|}{10.7} 			& \multicolumn{2}{c|}{6.7 } 		\\ \specialrule{1.5pt}{0pt}{0pt} 
			& 		Pol. 	& $n$ 		& $b$ 		& $f_0$ [GHz]	& XPD [dB]	& $\sigma$ [dB]						\\ \hline
			CIF 	& V-V 	& 2.8 		& 0.21 		& \textcolor{black}{49}  			& - 		& 7.5 								\\ \hline
			CIFX 	& V-H 	& 2.8 		& 0.21 		& \textcolor{black}{49}  			& 10.6  	& 5.5  								\\ \specialrule{1.5pt}{0pt}{0pt} 
			& 		Pol. 	& $\alpha$ 	& $\beta$ 	& $\gamma$ 		& XPD [dB] 	& $\sigma$ [dB]						\\ \hline
			ABG 	& V-V 	& 4.1 		& -12.2 	& 3.8  			& - 		& 6.4 	 							\\ \hline
			ABGX 	& V-H 	& 4.1 		& -12.2 	& 3.8  & 12.3 dB & 5.5 dB \\ \specialrule{1.5pt}{0pt}{0pt}
			
			\multicolumn{7}{|c|}{\textbf{28 GHz and 73 GHz Multi-Frequency Omnidirectional}} 						\\
			\multicolumn{7}{|c|}{\textbf{NLOS Closed-Plan Path Loss Model Parameters}} 								\\ \specialrule{1.5pt}{0pt}{0pt}
			&    	Pol. 	& PLE 		& \multicolumn{2}{c|}{XPD [dB]} 		& \multicolumn{2}{c|}{$\sigma$ [dB]}\\ \hline
			CI 		& V-V 	& 3.0 		& \multicolumn{2}{c|}{-} 				& \multicolumn{2}{c|}{11.4 } 		\\ \hline
			CIX 	& V-H 	& 3.0 		& \multicolumn{2}{c|}{13.4 } 			& \multicolumn{2}{c|}{11.2 } 		\\ \specialrule{1.5pt}{0pt}{0pt} 
			& 		Pol. 	& $n$ 		& $b$	 	& $f_0$ [GHz]	& XPD [dB]	& $\sigma$ [dB]						\\ \hline
			CIF 	& V-V 	& 3.0 		& 0.20 		& \textcolor{black}{50}  			& - 		& 10.9 								\\ \hline
			CIFX 	& V-H 	& 3.0 		& 0.20 		& \textcolor{black}{50} 			& 13.5 	 	& 10.1 	 							\\ \specialrule{1.5pt}{0pt}{0pt} 
			& 		Pol. 	& $\alpha$ 	& $\beta$ 	& $\gamma$ 		& XPD [dB]	&$\sigma$ [dB]						\\ \hline
			ABG 	& V-V 	& 2.8 		& 6.2 		& 3.8  			& - 		& 10.8 								\\ \hline
			ABGX 	& V-H 	& 2.8 		& 6.2 		& 3.8  			& 13.3  	& 9.8 							 	\\ \specialrule{1.5pt}{0pt}{0pt}
		\end{tabular}}
	\end{table}
\subsection{Multi-Frequency Models}

\subsubsection{Multi-Frequency Models for Co- and Cross-Polarized Antennas}
\begin{figure}
	\centering
	\includegraphics[width=3.7in]{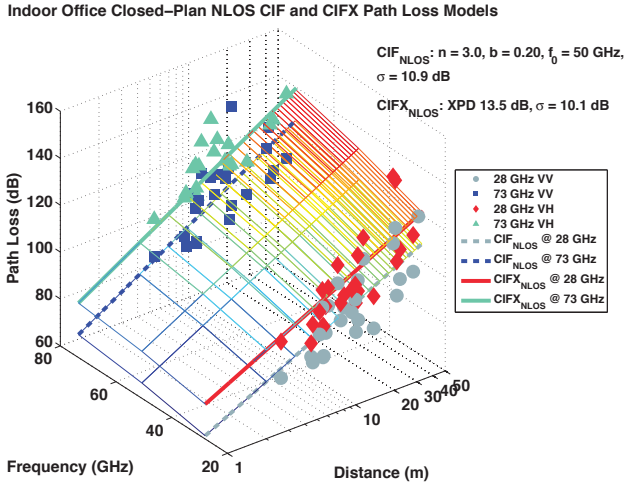}
	\caption{28 GHz and 73 GHz multi-frequency omnidirectional CIF and CIFX model scatter plots and parameters for the closed-plan indoor NLOS environment and layout.}
	\label{fig:MFCIF}
\end{figure}

Table~\ref{tbl:OmniMFPL} provides the 28 GHz and 73 GHz omnidirectional multi-frequency CI, CIX, CIF, CIFX, ABG, and ABGX path loss models for all environments and indoor layouts. Similar to single frequency models, the multi-frequency model XPDs (given in Table~\ref{tbl:OmniMFPL}) in LOS are larger than in NLOS, indicating more significant de-polarized effects due to penetration, reflection, and diffraction from obstructions. The three-parameter ABG models for all environments and layouts resulted in lower SF standard deviations than the one-parameter CI model, while the difference in standard deviations in a majority of cases is within 1 dB (the largest difference is 1.6 dB for the NLOS open-plan layout), and by introducing a frequency dependent term to create the two-parameter CIF model, SF standard deviation differences for CIF and ABG models are smaller (within 0.9 dB for all environments and layouts, less than an order of magnitude smaller than the standard deviation of the NLOS closed-plan models). Fig.~\ref{fig:MFCIF} shows the multi-frequency CIF and CIFX path loss models for the NLOS closed-plan layout. The gap in the CIF and CIFX model planes in Fig. 4 is exactly 13.5 dB (XPD factor) between the two models at any frequency at any distance for co- vs. cross-polarized antennas.  
	
\subsubsection{Multi-Frequency Models for Combined Antenna Polarization}
All co- and cross-polarized omnidirectional measurement data at 28 GHz and 73 GHz were combined and used to develop the combined polarization CI, CIF, and ABG omnidirectional multi-frequency path loss models as provided in Table~\ref{tbl:OmniMFPLcomb}. The CIF models for the combined polarization scenario result in lower SF standard deviations than the CI models, but differences are 0.9 dB or lower between each environment and layout, due to the frequency balancing parameter $b$ in the CIF model. The three parameter ABG model has the lowest SF standard deviation in all cases, but by no more than 1.1 dB and 0.4 dB compared to the CI and CIF models, respectively, in all environments and layouts (very little improvement, considering the typical SF standard deviations for all models are about 10 dB to 13 dB). 
\section{Conclusion}
This paper presented 28 GHz and 73 GHz single frequency and multi-frequency path loss models for combined, co-, and cross-polarized antenna combinations. The comparison of the CI and FI single frequency models resulted in similar performance regarding the SF standard deviations for all environments and layouts. The one-parameter CI model that includes a standardized 1 m anchoring point, allows for simple calculations (without losing accuracy), stable prediction of path loss beyond the measured region, and for straightforward comparisons across frequency bands, environments, and layouts. The CIF model with a frequency dependent term offers similar performance compared to the more complex ABG model that is not grounded by a close-in free space reference path loss. Furthermore, the large variations in the FI and ABG model parameters can lead to large errors when extrapolating the model outside of the measurement range or without a broad range of distances and frequencies to optimize the model parameters. It is reasonable to use a 1 meter reference distance even in NLOS environments, because the propagating signal will most likely not encounter obstructions or blockages in the first meter of propagation even if the receiver is in a NLOS location (still captures true propagation in the first meter). The solid physical basis in both frequency and distance in the CIF model motivates its use for modeling indoor mmWave communications systems, while the CI model is more suitable for outdoor systems where the path loss has less dependence on frequency~\cite{Shu:VTC2016}.

\begin{table}
	\centering
	\renewcommand{\arraystretch}{1.3}
	\caption{Combined polarization multi-frequency omnidirectional path loss model parameters for LOS and NLOS environments and layouts. ``Env." stand for environment, ``L/O." stands for layout. ``co" stands for corridor, ``op" stands for open-plan, and ``cp" stands for closed-plan.}\label{tbl:OmniMFPLcomb}
	\scalebox{0.95}{
		\fontsize{8}{8}\selectfont
		\begin{tabular}{|c|c|c|c|c|c|c|} \specialrule{1.5pt}{0pt}{0pt}
			\multicolumn{7}{|c|}{\textbf{28 GHz and 73 GHz Multi-Frequency Combined Polarization}} \\
			\multicolumn{7}{|c|}{\textbf{Omnidirectional Path Loss Model Parameters}} \\ \specialrule{1.5pt}{0pt}{0pt}
			Model & Env. & L/O & \multicolumn{2}{c|}{PLE} & \multicolumn{2}{c|}{$\sigma$ [dB]}\\ \hline
			\multirow{5}{*}{CI} & \multirow{2}{*}{LOS}	& co & \multicolumn{2}{c|}{2.0}  & \multicolumn{2}{c|}{10.6} 	\\ \cline{3-7}
			& 						& op & \multicolumn{2}{c|}{2.4}  & \multicolumn{2}{c|}{9.8} 	\\ \cline{2-7}
			& \multirow{3}{*}{NLOS}	& co & \multicolumn{2}{c|}{3.1}  & \multicolumn{2}{c|}{11.6} 	\\ \cline{3-7}
			& 						& op & \multicolumn{2}{c|}{3.2}  & \multicolumn{2}{c|}{9.3} 	\\ \cline{3-7}
			& 						& cp & \multicolumn{2}{c|}{3.6}  & \multicolumn{2}{c|}{13.3} 	\\ \specialrule{1.5pt}{0pt}{0pt}
			Model & Env. & L/O & $n$ & $b$ & $f_0$ [GHz] & $\sigma$ [dB] \\ \hline
			\multirow{5}{*}{CIF} & \multirow{2}{*}{LOS}	& co & 2.0 & 0.30 & 51  & 10.2  	\\ \cline{3-7}
			& 						& op & 2.4 & 0.36 & 51  & 9.2  		\\ \cline{2-7}
			& \multirow{3}{*}{NLOS}	& co & 3.1 & 0.23 & 51  & 10.7  	\\ \cline{3-7}
			& 						& op & 3.2 & 0.23 & 49  & 8.6  		\\ \cline{3-7}
			& 						& cp & 3.6 & 0.22 & 49  & 12.6  	\\ \specialrule{1.5pt}{0pt}{0pt}
			Model & Env. & L/O & $\alpha$ & $\beta$ & $\gamma$ & $\sigma$ [dB]\\ \hline
			\multirow{5}{*}{ABG} & \multirow{2}{*}{LOS}	& co & 0.7 & 22.7 & 3.5  & 9.8  	\\ \cline{3-7}
			& 						& op & 1.9 & 10.1 & 3.6  & 9.1  	\\ \cline{2-7}
			& \multirow{3}{*}{NLOS}	& co & 3.3 & \textcolor{black}{-7.1} & 4.2  & 10.6  	\\ \cline{3-7}
			& 						& op & 3.3 & -1.0 & 4.0  & 8.4 		\\ \cline{3-7}
			& 						& cp & 2.8 & 6.6  & 4.2  & 12.2  	\\ \specialrule{1.5pt}{0pt}{0pt}
		\end{tabular}}
	\end{table}
\bibliography{VTC_s2016_Indoor_Env_Path_Loss_v7_2}
\bibliographystyle{IEEEtran}
\end{document}